\documentclass[a4paper,20pt]{article}
    \usepackage[T1]{fontenc}
    \usepackage{graphicx}
    \usepackage{epsfig}
    \usepackage{amsmath}
    \usepackage{amsfonts}
    \renewcommand{\abstract}{}
\begin{document}
\makeatletter
\renewcommand{\@oddhead}{\textit{YSC'15 Proceedings of Contributed Papers} \hfil \textit{A.A. Stupka}}
\renewcommand{\@evenfoot}{\hfil \thepage \hfil}
\renewcommand{\@oddfoot}{\hfil \thepage \hfil}

\newcommand{\st}{\raisebox{.1 ex}{\mbox{$\,_{\stackrel{\hbox to 10 mm
{\rightarrowfill}} {\,t\gg\tau_{0}}}$}}}     

\fontsize{11}{11} \selectfont

\title{Reduced Description of Stellar Dynamics \\ by  Moments of Gravitation Field }
\author{\textsl{A.A. Stupka}}
\date{}
\maketitle
\begin{center} {\small
Dnipropetrovs'k National University,  Gagarin ave., 72, 49010
Dnipropetrovs'k, Ukraine\\antonstupka@mail.ru
 }
\end{center}

\begin{abstract}
Because of absence  of time derivatives from scalar potential  as
a generalized coordinate of  gravitation field (GF) in action of
nonrelativistic gravitating system, application of the Hamilton
method for description of  GF mechanics was impossible. In the
paper a transformation of the generalized coordinate of
 GF, that is based on continuity equation   and minimal action
principle, is proposed. A potential vector is introduced that is
similar to fixing of Hamilton gauge of the electromagnetic field.
This transformation  gives possibility of the calculation a
Hamilton function (HF), removes mathematical troubles of the Jeans
theory (Jeans swindle) and allows to construct kinetic theory of
GF using statistical mechanics methods.
\end{abstract}

\section*{Introduction}
\indent \indent Nowadays we have powerful computers that allow us
to calculate motion of every particle of complicated systems, but
for macroscopic systems like galaxies or clusters we still need to
simplify our description.

It is well known that main equations (kinetic or hydrodynamic) of
stellar dynamics theory use self-consistent GF \cite{pf76}. That
means, all correlation moments of GF are neglected and the first
moment (mean field) is considered as external field.

On the other hand, to construct more accurate theory we need HF of
masses with gravitation interaction. Such function has been
written many years ago by Hamilton himself, but without
generalized coordinates (and freedom degrees) of GF! That is
effective HF for masses only which uses special solution of
Poisson equation to deliver from GF.

We will try to use analogy with longitudinal electromagnetic field
\cite{3} to construct a transformation of generalized coordinates
of GF, which gives HF with GF coordinates. That will allow us to
use statistical mechanics methods to describe GF not only by mean
field, but also, for example, by one particle distribution
functions (normal and anomalous) which are equivalent to second
correlation moments of GF, that gives more information about GF
and whole system.

\section*{Mechanics}
\indent \indent We shall start from action for Newtonian GF  and
masses $m$ (equal for simplicity) with density $\sigma$
(\cite{ll2}\S 106)

\begin{equation}\label{a}
S = \int {\int {\left( {\frac{{m\sigma \upsilon ^2 }}{2} - m\sigma
\varphi  - \frac{{\left( {\nabla \varphi } \right)^2 }}{{8\pi G}}}
\right)} } dVdt
\end{equation}

 It is convenient to construct perturbation
theory by interaction when we have "small charge". To this purpose
make next generalized coordinate transformation
$\varphi=\sqrt{G}\varphi$.
\begin{equation}\label{an}
S = \int {\int {\left( {\frac{{m\sigma  \upsilon ^2 }}{2} -
\sqrt{G}m\sigma  \varphi  - \frac{{\left( {\nabla \varphi }
\right)^2 }}{{8\pi }}} \right)} } dVdt
\end{equation}

And we introduce an often used gravitational charge of particle
$a$: $e_{a}=\sqrt{G} m_{a},$ that gives
\begin{equation}\label{akt}
S = \int {\int {\left( {\frac{{m\sigma  \upsilon ^2 }}{2} -
e\sigma  \varphi  -\frac{{\left( {\nabla \varphi } \right)^2
}}{{8\pi }}} \right)} } dVdt
\end{equation}
This action looks like action for longitudinal electric field in
fixed Coulomb gauge (with minus, of course). It is well known,
that (\ref{akt}) gives us only effective HF, because there are no
time derivatives from scalar potential of GF. That means, scalar
potential is not a convenient generalized coordinate and we need
transform it.

Obviously, we have the continuity equation  (from charge
conservation fact)
\begin{equation}\label{4a} div\vec j + \partial _t \rho = 0,
\end{equation}
(density and current of mass are respectively
 $\rho  =\sum_{a} e_{a}\delta \left(
{x - x_a } \right)$ and $\vec j =\sum_{a} e_{a}\vec \upsilon
\delta \left( {x - x_a } \right)$). This is a common physical
fact, but
 we can not obtain it with Noether
theorem from the action (\ref{akt}). However, we can make
 gauge transformation using (\ref{4a}) in standard way. We
 introduce a new function of time and coordinates $\lambda$ and, after
 integration by parts, obtain
\begin{equation}\label{gauge}
  \int {\int {\left( {div\vec j + \partial _t \rho} \right)\lambda} }
  dVdt=-\int {\int {\left( {\vec j \nabla\lambda+ \rho\partial _t \lambda } \right)} }
  dVdt
\end{equation}
and subtract (\ref{gauge}) form (\ref{akt}) to obtain combination
$ \int {\int {\rho\left( {\partial _t \lambda -\varphi} \right)} }
dVdt $ that allows to vanish $\varphi$ in such a way
\begin{equation}\label{g}\varphi  =
\partial _t \lambda.
\end{equation}

Then let introduce a potential $ \vec A =c \nabla \lambda,$ that
will be a new (dynamical) coordinate of GF ($c$ is an arbitrary
constant, introduced for analogy with electric field). Now, after
using (\ref{g}) in last term (\ref{akt}), we obtain that the
action is
\begin{equation}\label{v0} S = \int {\int {\left( {\frac{{\rho
\upsilon ^2 }}{2} + \vec j\vec {{A}}/c -\frac{{\left( {\partial _t
\vec {{A}}/c} \right)^2 }}{{8\pi }}} \right)} } dVdt
\end{equation}
and we have generalized velocity proper to the new coordinate. We
now introduce a notation for GF strength $ -c\vec E = $ $=
\frac{{\partial \vec A}}{{\partial t}}. $ From (\ref{v0}) we take
a
  Lagrange  function
\begin{equation}\label{lagr}
L = L_m  + L_f  = \int {d^3x\left( {\frac{{\rho \upsilon ^2 }}{2}
+  \frac{\vec j\vec A}{c} -\frac{{\vec E^2 }}{{8\pi }}} \right)}
\end{equation}

Accordingly to standard procedure a generalized momentum is $P_n
\left( {x,t} \right) = \frac{1 }{{4\pi c}}E_n \left( {x,t}
\right)$ and we construct the nonrelativistic HF of masses and GF
$ \hat H = \hat H_{\rm m}  + \hat H_{\rm f}+ {\hat H_1 + \hat H_2
} , $ where terms with GF are
\begin{equation}
\hat H_{\rm f} =  - \int {d^3 x\frac{{\hat {\vec E}(x)^2 }}{{8\pi
}}},\,\hat H_1 = -\frac{1}{c} \int {d^3 x} \hat A_n (x)\hat j_{on}
(x), \,\hat H_2 = {\textstyle{1 \over 8\pi c^{2}}}\int {d^3 x\hat
A^2 (x} )\hat \Omega^{2} (x),\label{5}
\end{equation}
and $\Omega=\sqrt{4\pi \rho}$ is Jeans frequency, $\hat H_{\rm m}$
is HF of free masses (without nonrelativistic GF).

\section*{Dynamical equations for GF moments}
\indent \indent
A system with macroscopic number of particles
(stars) can be described by distribution function which satisfy
Liouville equation $
\partial _t \rho (t) =  \left\{ {\hat H,\rho \left( t \right)} \right\} . $ Then dynamical (time) equation
for arbitrary physical value can
be written as $\dot A_n \left( {x,t} \right) =-Sp\rho \left( t
\right)\left\{ {\hat H,\hat A_n \left( x \right)} \right\}.$ For
potential and strength of GF we have
\begin{equation} \dot {\vec A }=  - c\,\vec E
,\,\label{27} \dot {\vec E} =  4\pi \vec J
\end{equation}
 where average current of mass
$
  \vec J\left( {x,t} \right)= Sp\rho \left( {t
} \right)\hat{ \vec J}\left( x \right) $ has such microscopic
structure $ \hat J_n \left( x \right) \equiv \hat j_n \left( x
\right) - $ $-\frac{1}{4\pi c} \hat A_n \left( x \right)\hat
\Omega \left( x \right). $ And we can write in such a manner
equations for any moments of GF. Let us consider GF kinetic theory
approximation, that means we must construct equations for the
second space correlation moments of  GF  \cite{3,5}
\begin{equation}
\nonumber \frac{{\partial \left( {\hat A_n\hat A'_m }
\right)}}{{\partial t}} =  - c\,\left( {\hat E_n\hat A'_m }
\right) - c\left( {\hat A_n\hat E'_m } \right), \label{45b}
\frac{{\partial \left( {\hat E_n\hat A'_m } \right)}}{{\partial
t}} =  - c\left( {\hat E_n\hat E'_m } \right) + 4\pi \left( {\hat
J_n\hat A'_m } \right)
\end{equation}
\begin{equation}\label{47b}
\frac{{\partial \left( {\hat A_n\hat E'_m } \right)}}{{\partial
t}} =  - c\left( {\hat E_n\hat E'_m } \right) + 4\pi \left( {\hat
A_n\hat J'_m } \right), \frac{{\partial \left( {\hat E_n\hat E'_m
} \right)}}{{\partial t}}= 4\pi  \left( {\hat J_n\hat E'_m }
\right) + 4\pi \left( {\hat E_n\hat J'_m } \right)
\end{equation}
Sources are correlations between subsystems of GF and particles $
  Sp\rho \left( t \right) {\hat J_n(x)\hat A_m(x') }  =
\left( {J_n A'_m } \right). $

Gravitation interaction is weak and then we can use Bogolyubov
boundary condition of complete correlation weakening \cite{4} to
find solution of Liouville equation in perturbation theory. We can
divide HF (\ref{5}) into main part and weak interaction. GF main
part of HF is
\begin{equation}
\hat H_{f}  =   -\int {d^3 x\frac{{\hat {\vec E}(x)^2 }}{{8\pi
}}}+ {\textstyle{1 \over 8\pi c^{2}}}\int {d^3 x\int {d^3 x'\hat
{\vec{A}} (x)\hat {\vec{A}} (x')\omega^{2} (x-x')}},\label{52}
\end{equation}
where the function $\omega^{2} (x-x')$ will be obtained as a
solution of dispersion equation of linear part of time equations.

Let suppose for simplicity that subsystem of free particles is
thermostat. Then we obtain with the help of reduced description
method a distribution function in the first order of
nonrelativistic interaction  (see \cite{3})

\begin{equation}\label{30}
\rho \left( t \right) = \rho _f \left( t \right)w -\frac{1}{c}
\int\limits_{ - \infty }^0 {d\tau \int {d^3x\left\{ {\hat A_n
\left( {x,\tau } \right)\hat j_n \left( {x,\tau } \right),\rho _f
\left( t \right)w} \right\} } }.
\end{equation}
Fourier-component of GF potential in interaction picture is $ \hat
A_{nk}^l \left( \tau  \right) = ch(\omega_{k}\tau)\hat A_{nk}^l  +
\frac{ c}{\omega_{k}}sh(\omega_{k}\tau) \hat E_{nk}^l, $ where
longitudinal vector is $\hat A_{nk}^l \equiv \hat A_{mk}\tilde k_m
\tilde k_n $, $\tilde k_n \equiv \frac{{k_n }}{k}$.

 \section*{Kinetic coefficients in the GF equations }
\indent \indent Averaging with distribution function  (\ref{30})
sources in (\ref{27}) and (\ref{47b}), we obtain current up to the
second order, which is linear on GF parameters
\begin{equation}\label{36}
J_n \left( {x,t} \right) = \int d^3x'\sigma _{nl} \left( {x - x'}
\right)E_l \left( {x',t} \right) +\int {d^3x'\lambda _{nl} \left(
{x - x'} \right)A_l \left( {x',t} \right)} ,
\end{equation}
where kinetic coefficients of second order for GF are
\begin{equation}\label{37}
\sigma _{nl,k}  = -\left(G^l \left( {k,-i\omega} \right) - G^l
\left( {k,i\omega} \right)\right)\tilde k_n \tilde k_l
/2\omega,\,\lambda _{nl,k}  = -\left( { \Omega^{2}/2\pi  +  G^l
\left( {k,i\omega} \right)+G^l \left( {k,-i\omega} \right)}
\right)\tilde k_n \tilde k_l/2c .
\end{equation}
Here we have used a Fourier-transformed Green function
\begin{equation}\label{34}
  G_{nl}\left( {x,t} \right) =  \theta \left( t
\right)Sp w\left\{ {\hat j_n \left( {x,t} \right),\hat j_l \left(
0 \right)} \right\}  = \int {\frac{{d^3 kd\omega }}{{(2\pi )^4
}}G_{nl}\left( {\vec k,\omega } \right)} e^{i\vec k\vec x -
і\omega t}
\end{equation}
for homogeneous and isotropic medium $G_{nl} \left( {\vec k,\omega
} \right) = G^t \left( {k,\omega } \right)\delta _{nl}^t + G^l
\left( {k,\omega } \right)\tilde k_n \tilde k_l .$

For average GF-particles correlations we obtain
\begin{eqnarray}
 \nonumber (A_n^x J_l^{x'})^t  = \int d^3x''\left\{ {\sigma_{lm} \left( {x' -
x''} \right)(A_n^x E_m^{x''})^t }\right. \left.{+\lambda _{lm}
\left( {x' - x''} \right)(A_n^x A_m^{x''})^t } \right\} +
S_{nl}\left( {x - x'} \right) \\\label{51b}
 (E_n^x J_l^{x'})^t  = \int d^3x''\left\{ {\sigma_{lm}
\left( {x' - x''} \right)(E_n^x E_m^{x''})^t  }\right. \left.{
+\lambda _{lm} \left( {x' - x''} \right)(E_n^x A_m^{x''})^t }
\right\} + T_{nl}\left( {x - x'} \right).
\end{eqnarray}
where free terms look like Langevine force correlations of
phenomenological theories
\begin{equation}\label{53b}
S_{nl}( k)=- 4\pi T c^{2}\lambda_{nl} ( k)/\omega^{2}_k, \quad
T_{nl}(k) = 4\pi T \sigma_{nl}( k).
\end{equation}

\section*{Solutions of the GF equations }
\indent \indent We shall search a solution  as
$A(t)=Ae^{\omega_{k} t}.$ Then from (\ref{27}) with (\ref{36}) we
obtain
\begin{equation}\label{38}
\omega_{k} A_{nk}^l  =  -c E_{nk}^l,\,\omega_{k} E_{nk}^l  = -4\pi
\epsilon\left( {\sigma _k E_{nk}^l  + \lambda _k A_{nk}^l }
\right)
\end{equation}
Dispersion equation for (\ref{38}) is $ \omega_{k}  = 4\pi \left(
{\sigma _k - \lambda _k \frac{c}{\omega_{k}} } \right) $ or after
using (\ref{37}) $ \omega^{2}_{k}  =  \Omega^{2}+4\pi  G^l \left(
{k,i\omega_{k}} \right) .$ It gives well known results for example
with Maxwell distribution function of particles. For big wave
vector and one type of particles we have $\omega_k=-\frac{1  -
\frac{1}{{r^{2}_{D}k^{2} } }}{\sqrt{\pi/2}}
r^{2}_{D}\upsilon_{T}k^{3}, $ where $\upsilon^2_{T}=T/m$ and
$r_{D}=\upsilon_{T}/\Omega$. For small wave vector we have
increment $ \omega_k=\pm \sqrt{\Omega^{2}-3 k^{2}\upsilon^2_{T}}.
$

When we have external current with increment $\omega$, we must
change in (\ref{52}) $\omega^{2} (x-x')\rightarrow\omega^{2} $ and
go on to obtain $ \omega A_{nk}^l  =  -c E_{nk}^l,\,\omega
E_{nk}^l  = 4\pi \left( {\sigma _k E_{nk}^l  + \lambda _k
A_{nk}^l+j_{nk}^{l ext}} \right) .$ If we return to standard
scalar potential $E=-\nabla\varphi$ then for point mass and zero
frequency we obtain gravitation screening
$
    \varphi(\vec{r})=-\sqrt{G}m^{ext}\frac{cos(\frac{r}{r_{D}})}{r}.
$

It is well known that Jeans theory  in equilibrium state requires
$\varphi=0$ (\cite{zn75} p.273), but Poisson equation forbids it
(Jeans swindle). Using (\ref{27}) in hydrodynamical medium we have
equation $ {\dot{ E}_{n} }=   \,4\pi \rho u_n $ that has zero
equilibrium solution.

But in equilibrium state  GF is not absent. It has nonzero values
of second correlation moments. For example, energy of GF is
proportional to $(E_n E_m({x'-x}))$ correlation moment and from
(\ref{47b}) we obtain $
  (EE)^k  =  - 4\pi T ,
$ where $T$ is thermostat temperature.

Solution for linear equations (\ref{45b}) with (\ref{51b}) has the
form of combination frequencies $ A(t)=Ae^{(\omega_{k}+\omega_{k'}
)t}. $ For homogeneous and isotropic but nonstationary GF, when
the first moments are zero, for the second moments we obtain time
dependence with $2\omega_k .$

\section*{Results and Conclusions}
\indent \indent So, we have transformed nonrelativistic action for
masses and GF  using continuity equation and introduced potential
vector as a new generalized coordinate of GF. That allows us to
obtain real (not effective!) HF for this system with generalized
coordinate and momentum of GF. Using Bogolyubov reduced
description method with special GF main HF we have obtained time
equations for the first and the second correlation moments of GF,
that are linear in perturbation theory of weak gravitation
interaction.
 Equations for mean GF give  zero equilibrium solution,
unlike the Poisson equation, that removes mathematical troubles of
the Jeans theory (Jeans swindle). Expressions for kinetic
coefficients in terms of the Green function of the currents are
constructed. Solutions for  GF equations in the case of
 Maxwell  particles distribution    are found.

\section*{Acknowledgement}
\indent \indent I would like to thank Prof. A. I. Sokolovsky and
Dr. E. M. Kopteva for helpful discussions. This work was supported
by the State Foundation for Fundamental Research of Ukraine
(project No. 25.2/102).


\end{document}